\documentclass[aps,twocolumn]{revtex4-1}
\usepackage{adjustbox}
\usepackage{epsfig}
\usepackage{graphicx}
\usepackage{amsmath,amssymb,amsfonts}
\usepackage{array}
\usepackage{url}
\usepackage[colorlinks=true,linktocpage=true,linkcolor=blue,citecolor=blue]{hyperref}
\usepackage{multirow}
\usepackage{float}
\usepackage{xspace}
\usepackage{subcaption}
\usepackage{booktabs}
\usepackage[usenames,dvipsnames]{color}
\usepackage{hyperref}
\usepackage{xspace}
\usepackage{multirow}
\usepackage{ragged2e}

\newcommand{\sqs}{\mbox{$\sqrt{s} =$}\xspace}
\newcommand{\pt}{$p_{\rm{T}}$\xspace}

\newcommand{\nch}{$N_{\rm ch}$\xspace}

\newcommand{\inj}{$``$in-jet''\xspace}
\newcommand{\oj}{$``$out-of-jet''\xspace}
\newcommand{\jty}{$``$jetty''\xspace}
\newcommand{\iso}{$``$isotropic''\xspace}
\newcommand{\sph}{$S_\mathrm{O}$\xspace}

\begin{document}

\title{Searching for enhancement in coalescence of in-jet (anti-)deuterons in proton-proton collisions}
\author{Yoshini Bailung$^{1}$} \thanks{yoshini.bailung.1@gmail.com}
\author{Neha Shah$^{2}$}
\thanks{nehashah@iitp.ac.in}
\author{Ankhi Roy$^{1}$}
\thanks{ankhi@iiti.ac.in}
\affiliation{$^{1}$Department of Physics, Indian Institute of Technology Indore, Simrol, Indore, 453552, Madhya Pradesh, India}
\affiliation{$^{2}$Department of Physics, Indian Institute of Technology Patna, Bihta, Patna, 801106, Bihar, India}

\begin{abstract}

Recent measurements from ALICE report that $``$in-jet'' nucleons carry a higher probability of forming a deuteron via coalescence than the nucleons from the underlying event (UE). This study makes use of an event shape classifier to separate the $``$in-jet'' deuterons and the deuterons in the UE produced in high multiplicity proton-proton collisions at $\sqrt{s} = 13$ TeV. Event shape variables such as transverse spherocity allow the categorization of hard and soft components of an event, which can be divided into two respective classes; $``$jetty'' and $``$isotropic''. The $``$jetty'' deuterons minus the contribution of the deuterons from the $``$isotropic'' event are taken as $``$in-jet'' deuterons, and the coalescence mechanism is tested. The coalescence is performed with a Wigner function formalism, augmented as an afterburner to \textsc{pythia}8. The possible enhancement of the coalescence probability of $``$in-jet'' deuterons is investigated by calculating the coalescence parameter ($B_{2}$) in different spherocity classes in high-multiplicity $pp$ collisions. 

\end{abstract}
\date{\today}
\maketitle

\section{Introduction}
\label{introduction}
Experiments at facilities such as the CERN Large Hadron Collider (LHC) and BNL Relativistic Heavy Ion Collider (RHIC) have observed a handful of the light (anti-)nuclei and (anti-)hyper nuclei states produced in high energy heavy-ion and hadronic collisions~\cite{British-Scandinavian-MIT:1977tan, Alper:1973my, E878:1998vna, E802:1999hit, STAR:2016ydv, STAR:2019sjh, STAR:2010gyg, ALICE:2021mfm, ALICE:2015wav, ALICE:2017xrp, ALICE:2020foi, ALICE:2017jmf, STAR:2010gyg}. Investigations in the production and dynamics of these nuclear clusters are important to gain insight into the low energy quantum chromodynamics (QCD) interactions and also provide constraints on dark matter detection and baryon asymmetry of the universe~\cite{STAR:2010gyg, Winkler:2020ltd, ALICE:2022zuz}. The mechanism behind light-nuclei production, however, is still debatable. Popular microscopic models, based on the coalescence of nucleons at kinetic freeze-out or statistical hadronization models based on ensemble formalism at chemical freeze-out, are capable of describing the experimental measurements up to certain extents~\cite{Butler:1963pp, Kapusta:1980zz, Scheibl:1998tk, Zhao:2018lyf, Sun:2018mqq, Steinheimer:2012tb, Andronic:2010qu, Becattini:2014hla, Vovchenko:2018fiy, Kachelriess:2023jis}. Newer results from these experiments provide a wealth of information to test these models and develop a cohesive description of the nature of light-nuclei production mechanism.\\

A straightforward model to describe light-nuclei formation is via $``$coalescence'' of (anti-)nucleons. (Anti-)Nucleons close together in phase space can form (an)a (anti-)nuclei. In this approach, the invariant yields of light (anti-)nuclei and (anti-)proton at midrapidity are related via
\begin{equation}
   E_{A}\frac{d^3 N_{A}}{dp_{A}^3} = B_{A}  \left(E_{p}\frac{d^3N_{p}}{dp_{p}^3} \right)^{A},
   \label{eqnb2}
\end{equation}
where, $B_{A}$ is the coalescence parameter, and the (anti-)proton and (anti-)nuclei momentum are related as $p_{A} = A \cdot p_{p}$. The coalescence parameter quantifies the coalescence probability of the (anti-)nucleon pair and is inversely proportional to the emission source volume. In small systems, $B_{A}$ shows an inverse trend with the event multiplicity.\\

A recent measurement by ALICE provided some interesting ground on the likelihood of (anti-)deuteron production in jets~\cite{ALICE:2022ugx}. With the help of event topology variables, the deuteron was observed to have a higher chance of production via coalescence in jets. Another work on jet-associated deuteron production was done by ALICE, where the higher population of deuterons around a jet was observed via angular correlations of deuterons and charged particles~\cite{ALICE:2020hjy}. Due to the collimated nature of jets, the \inj hadrons produced from the fragmentation processes are strongly correlated in phase space. As a result, the hadrons that are close in space are also close in momenta. This correlation is absent for \oj or the UE hadrons. In a coalescence scenario, the \inj nucleons with phase space vicinity will have a larger chance to form a deuteron and, therefore, carry a higher probability than \oj nucleons.\\

With \textsc{pythia}8 and a simple coalescence afterburner, one can interpret these measurements in proton-proton ($pp$) collisions and have a naive understanding of the underlying mechanism behind coalescence.~\cite{ALICE:2022ugx, ALICE:2020hjy, ALICE:2017qfj}. The simple coalescence model used in the ALICE measurements uses a condition only on the relative momentum of the (anti-)nucleon pair ($\Delta p$)~\cite{ALICE:2017qfj}. However, to dive deeper into the microscopic aspects, incorporating an emission source with a probabilistic view of the structure of the deuteron is important. More advancements in coalescence now allow a probabilistic selection of the (anti-)nucleon pairs based on the Wigner density of the deuteron~\cite{JETSCAPE:2022cob, Kachelriess:2019taq, Kachelriess:2020amp}. Moreover, one can introduce spatial degrees of freedom in \textsc{pythia}8 by utilizing realistic emission sources extracted via femtoscopic correlations of baryon pairs in $pp$ collisions~\cite{ALICE:2020ibs, Mahlein:2023fmx}. A myriad of advanced coalescence models inspires this work to study (anti-)deuterons production in jets using event shape variables.\\

This work explores \jty and \iso deuteron production in $pp$ collisions at midrapidity using an advanced Wigner coalescence model. To classify the events based on their jet/event topology, transverse spherocity (\sph) is used. \sph is an event shape variable that tells us about the $``$jettiness'' of an event, which helps categorize the event as $``$hard'' or $``$soft''. With this in mind, we perform a multi-differential measurement of deuteron production with \sph and investigate the likelihood of deuterons in jets. The model runs as an afterburner and uses a Wigner formalism of coalescence, where the probability density is based on the deuteron wave function. The model also implements a particle emission conformity where relative distances between the nucleon pairs are parameterized with measurements from ALICE.\\

The paper is arranged in the following manner; Sec.~\ref{method} describes the working of the model that is factorized into four subsections. Each subsection describes an important aspect of the model starting from event generation with \textsc{pythia}8 (Sec.~\ref{pythia}), the afterburner with Wigner formalism of coalescence for producing deuterons (Sec.~\ref{themodel}), and the emission source model to introduce spatial correlations between the nucleon pairs  (Sec.~\ref{emissionsource}). A brief discussion on the event shape variable of choice is also added in Sec.~\ref{spherocity}. Sec.~\ref{results} presents a detailed discussion of the results and the important findings of this study. The paper is concluded in Sec.~\ref{conclusions}, mentioning the possibility of new insight with developments in the deuteron wave function and precise measurements for source size estimation in the future.

\section{Methodology}\label{method}

\subsection{Event generation}\label{pythia}
\textsc{pythia} is a well-known QCD inspired event generator that can be applied to a large set of phenomenological problems in high-energy as well as astroparticle physics~\cite{Sjostrand:2000wi,Sjostrand:2007gs}. \textsc{pythia}8 provides a wide range of processes and a variety of control parameters suitable for generating hard and soft scatterings, initial and final state radiations in parton scatterings, parton fragmentations, multi-parton interactions, and color-reconnection mechanisms in hadronization.\\

In this work, we use \textsc{pythia} v8.37, with multi-parton interactions (MPI) and color reconnection (CR) mechanisms at hadronization~\cite{Bierlich:2022pfr,Skands:2014pea,Sjostrand:1987su,Argyropoulos:2014zoa}. MPI enhances particle production in \textsc{pythia} by including more than one partonic scattering. Four to ten partonic interactions are expected in a single LHC event, depending on the overlap between the colliding proton beams~\cite{Sjostrand:2004pf}. MPI is essential to describe charged-particle production, the underlying event (UE) and to study the interplay between soft and hard scatterings in particle production. \\

Hadronization in \textsc{pythia}8 is carried out via the Lund string fragmentation model, where color strings between the partons and beam remnants are made to move apart~\cite{Andersson:1983ia}. New quark-antiquark pairs are formed when the strings break, and the process continues until small segments of the strings remain. This is the final step of hadronization, where the small pieces of strings are identified as hadrons. Before this, the CR mechanism can be implemented, rearranging the strings between the partons~\cite{Argyropoulos:2014zoa}. This is done by reducing the total length of the string, which in turn reduces the multiplicity of an event.\\

\begin{figure}
    \flushleft\includegraphics[scale = 0.45]{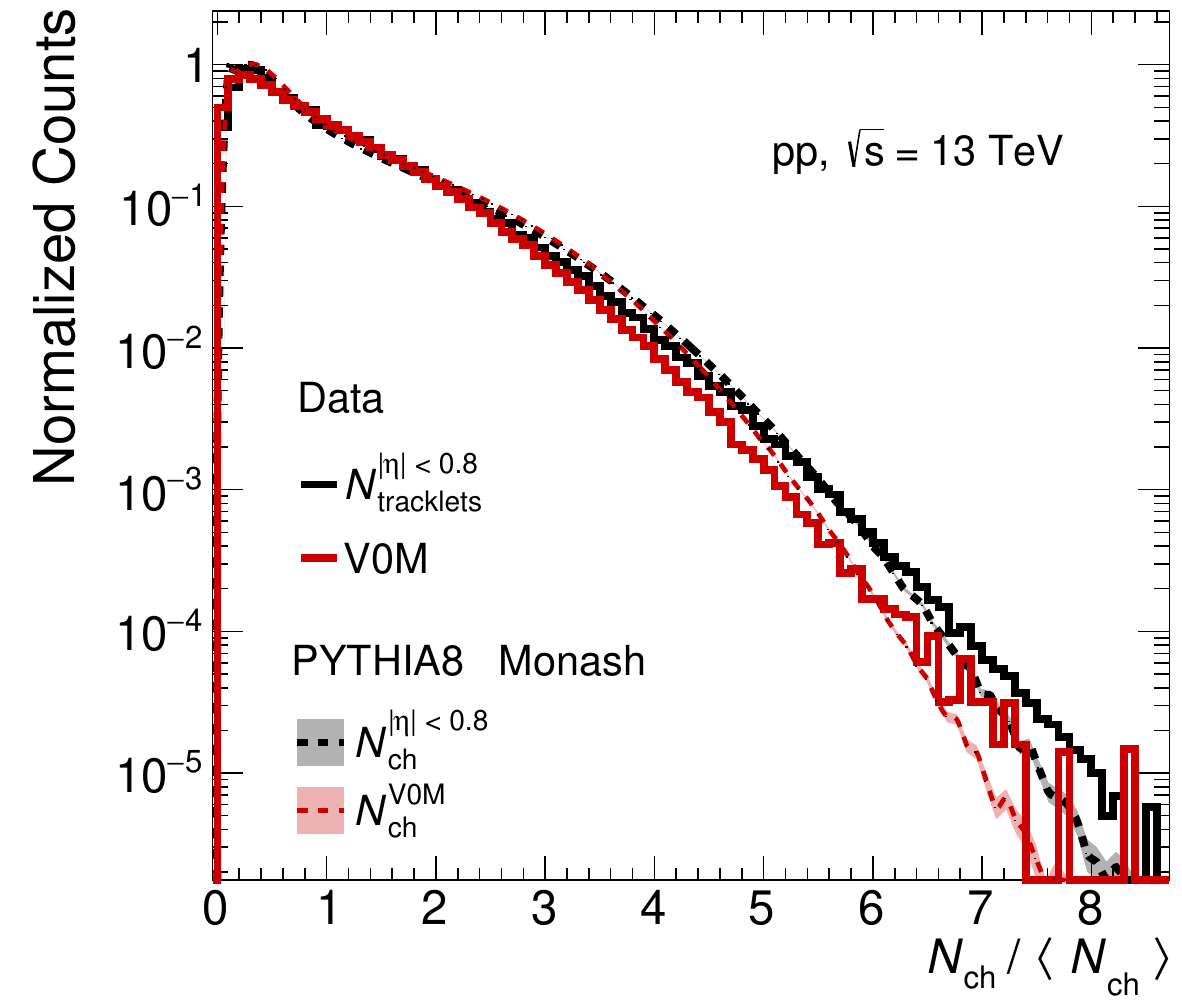}
    \caption{\justifying Charged-particle multiplicity distributions from \textsc{pythia}8 + Monash (dotted lines) for $pp$ collisions at \sqs 13 TeV. Data from ALICE (solid lines) for the midrapidity multiplicity estimator ($N_\mathrm{tracklet}^{|\eta|<0.8}$) and the V0M detector are presented for comparison~\cite{ALICE:2020swj, ALICE:2019avo}.}
    \label{fig:multdist}
\end{figure}

In Fig.~\ref{fig:multdist}, we report the results from \textsc{pythia}8 with the Monash 2013 tune~\cite{Skands:2014pea} for $pp$ collisions at \sqs 13 TeV, which uses a combined simulation of MPI and CR. ALICE measures charged-particle multiplicities with the number of track segments in two inner layers of the inner tracking system (ITS) at $|\eta|< 0.8$ $\left(N_\mathrm{ch}^{|\eta|<0.8}\right)$, and with the amplitudes from V0M scintillation detectors placed in the forward and backward region of the interaction vertex i.e. $-3.7<\eta<-1.7$ and $2.8<\eta<5.1$ $\left(N_\mathrm{ch}^{\rm{V0M}}\right)$. The charged-particle multiplicities (\nch) in $pp$ collisions at \sqs 13 TeV with \textsc{pythia}8 + Monash tune are calculated in the two acceptance ranges. For both acceptances, the distributions are presented by scaling the event multiplicity to the average multiplicity $(N_{\mathrm{ch}}/\langle N_{\mathrm{ch}}\rangle)$ which represents the fractional cross-section with the minimum-bias cross-section in $pp$ collisions. The \textsc{pythia}8 predictions reasonably describe the experimental results obtained from Refs.~\cite{ALICE:2020swj, ALICE:2019avo}.

\begin{figure}
    \centering
    \includegraphics[scale = 0.4]{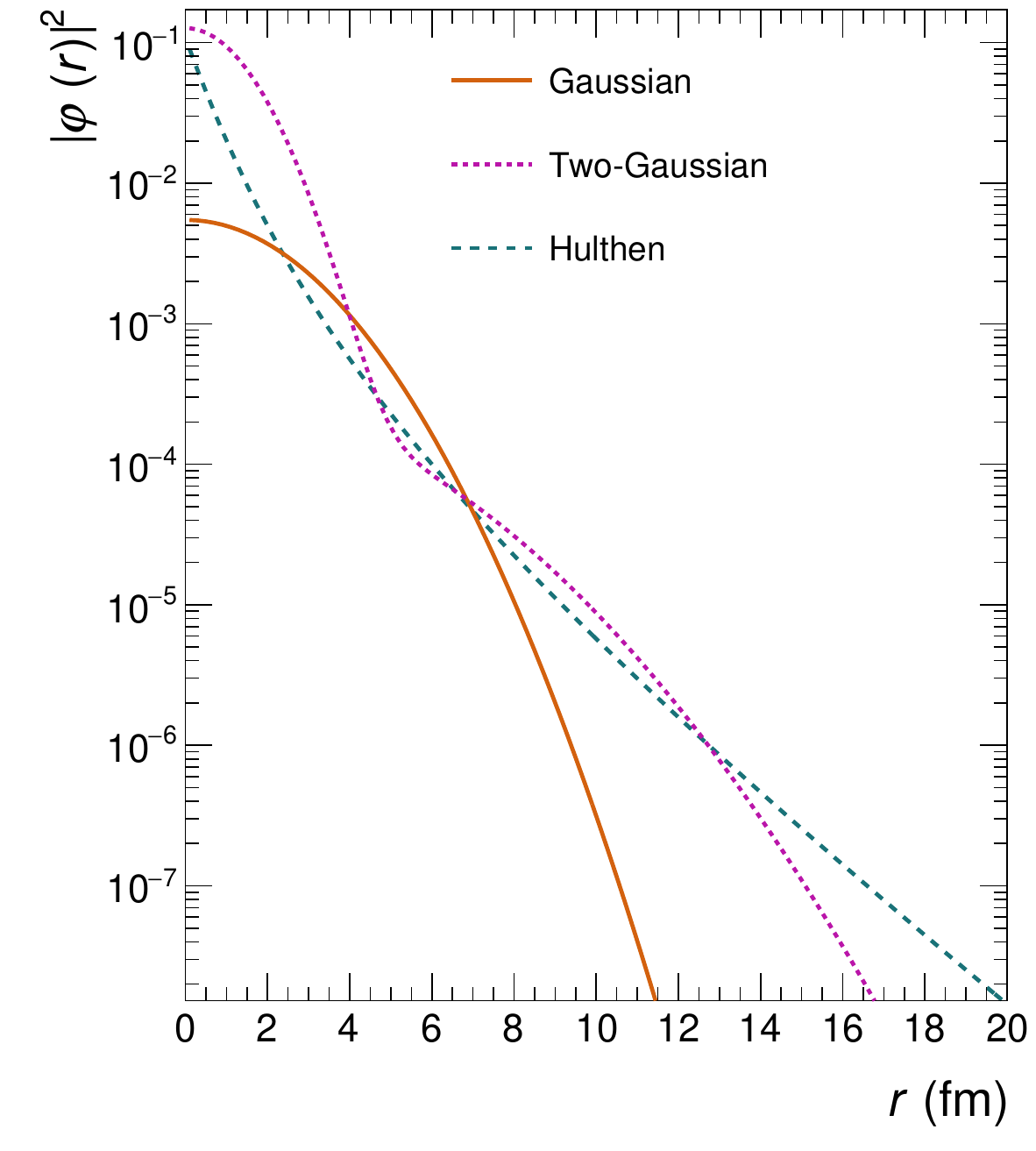}
    \caption{\justifying Wigner probability distributions of different deuteron wave functions.}
    \label{fig:wavefunc}
\end{figure}

\subsection{Wigner coalescence formalism}\label{themodel}

A quantum mechanical formalism of the deuteron coalescence process is presented in this section. The key ingredient here is the underlying wave function of the deuteron. In the rest frame of the deuteron, a nucleon pair having position coordinates $\mathbf{r}_{1}$, $\mathbf{r}_{2}$ and momenta $\mathbf{p}_{1}$, $\mathbf{p}_{2}$, is considered when they obey the relations
\begin{eqnarray}
    \mathbf{r}_{d} &=& \frac{1}{2}( \mathbf{r}_{1}+\mathbf{r}_{2}),\hspace{1cm}\mathbf{r} =  \mathbf{r}_{1} - \mathbf{r}_{2}\\
    \mathbf{p}_{d} &=&  \mathbf{p}_{1} + \mathbf{p}_{2},\hspace{1cm}\mathbf{q} =\frac{1}{2} (\mathbf{p}_{1}-\mathbf{p}_{2})
\end{eqnarray}
The invariant yield of the deuteron can be written into the form
\begin{eqnarray}
\label{maineqn}
  \frac{d^3N_{d}}{dp_{d}^3} = \mathcal{S}\int \frac{d^3r_{d}d^3rd^3q}{(2\pi)^6} \mathcal{D}(\mathbf{r}, \mathbf{q})\\ \nonumber
  \times W_{np}\left(\frac{\mathbf{p}_{d}}{2} + \mathbf{q}, \frac{\mathbf{p}_{d}}{2}-\mathbf{q},\mathbf{r}_{d} + \frac{\mathbf{r}}{2}, \mathbf{r}_{d} - \frac{\mathbf{r}}{2}\right)
\end{eqnarray}
    
where, $\mathcal{S} = 3/8$ is the statistical spin-isospin averaging factor, $W_{np}$ is the (anti-)nucleon pair selection probability term, and $\mathcal{D}$ is the deuteron Wigner function, given by
\begin{equation}
    \mathcal{D}(\mathbf{r}, \mathbf{q}) = \int d^3\xi e^{-i\mathbf{q}\cdot\xi}\varphi\left(\mathbf{r} + \frac{\mathbf{\xi}}{2}\right)\varphi^{*}\left(\mathbf{r} - \frac{\mathbf{\xi}}{2}\right)
\end{equation}
The $\varphi (r)$ is the choice of the deuteron wave function. In this work, two different forms of wave functions are chosen: a single Gaussian of the form
\begin{equation}
    \varphi (r) = \frac{1}{(\pi d^2)^{3/4}} e^{-\frac{r^2}{2d^2}}
\end{equation}
with $d$ = 3.2 fm~\cite{Bellini:2018epz}. This makes
\begin{equation}
    \mathcal{D}(\mathbf{r}, \mathbf{q}) = 8e^{-\frac{r^2}{d^2} - q^2 d^2}
\end{equation}
The other choice is a double or $``$two" Gaussian form parametrized to the Hulthen wave function for the deuteron. The formalism is adapted from the studies, which use double Gaussian wave functions that are parameterized to reproduce the ground-state deuteron~\cite{Kachelriess:2019taq, Kachelriess:2020amp, Kachelriess:2023jis}. The Hulthen wave function,
\begin{equation}
    \varphi (r) = \sqrt{\frac{ab(a+b)}{2\pi (a - b)^{2}}}\frac{e^{-ar} - e^{-br}}{r}
\end{equation}
with $a = 0.23$ fm$^{-1}$ and $b = 1.61$ fm$^{-1}$, is based on the Yukawa theory of interactions and provides a good description of the deuteron ground state wave function. The two-Gaussian wave function can be written as
\begin{equation}
    \varphi (r) = \frac{1}{\pi^{3/4}}\left[ \frac{\Delta^{1/2}}{d_1^{3/2}}e^{-\frac{r^2}{2d_{1}^2}} + e^{\iota \alpha} \frac{(1 - \Delta)^{1/2}}{d^{3/2}_{2}}e^{-\frac{r^2}{2d_{2}^2}}\right]
\end{equation}
Choosing $e^{\iota \alpha} = \iota$, the probability distribution becomes
\begin{equation}
    \mid\varphi (r)\mid^2= \frac{1}{\pi^{3/2}}\left[ \frac{\Delta}{d_1^{3}}e^{-\frac{r^2}{d_{1}^2}} + \frac{(1 - \Delta)}{d^{3}_{2}}e^{-\frac{r^2}{d_{2}^2}} \right]
    \label{fitfunc}
\end{equation}
The parameters $\Delta = 0.247$, $d_{1} = 5.343$ fm$^{-1}$, and $d_{2} = 1.81$ fm$^{-1}$ are taken from the references~\cite{Kachelriess:2019taq, Kachelriess:2020amp, Kachelriess:2023jis}, that are extracted by fitting Eq.~\ref{fitfunc} to the Hulthen wave function. The probability distributions $|\varphi (r)|^2$ are displayed in Fig.~\ref{fig:wavefunc}, which shows the one and two Gaussian distributions and the Hulthen probability distribution.\\

$W_{np}$ is probability term that selects the (anti-)proton and (anti-)neutron pair at positions $\mathbf{r}_{d} \pm \mathbf{r}/2$ and momenta $\mathbf{p}_{d}/2 \pm \mathbf{q}$. This term can be factorized into spatial and momentum components as
\begin{equation}
\label{wnpfactor}
    W_{np} = S(\mathbf{r_{p}}, \mathbf{r_{n}})\cdot Q(\mathbf{p}_{d}/2 + \mathbf{q},\mathbf{p}_{d}/2-\mathbf{q}).
\end{equation}
Assuming a Gaussian source and neglecting spatial correlations between proton and neutron, the spatial component can be written as 
\begin{equation}
    S(\mathbf{r}; r_{0}) = \frac{1}{(4\pi r_{0}^2)^{3/2}} e^{-\frac{r^2}{4r_{0}^2}}
    \label{src}
\end{equation}
where $r_{0}$ is the size of the nucleon pair emitting source or the source radius.\\

It should be noted that the single Gaussian wave function used in this study does not quantitatively reproduce the deuteron yields. The Gaussian form predicts a lower yield of deuterons compared to the experimental results. The results, here, are scaled twice to their actual values to visualize the yields better when compared to the experimental results. On the other hand, the two-Gaussian formalism predicts the experimental results quantitatively, providing a reasonable description.

\begin{figure}
    \centering
    \includegraphics[scale = 0.3]{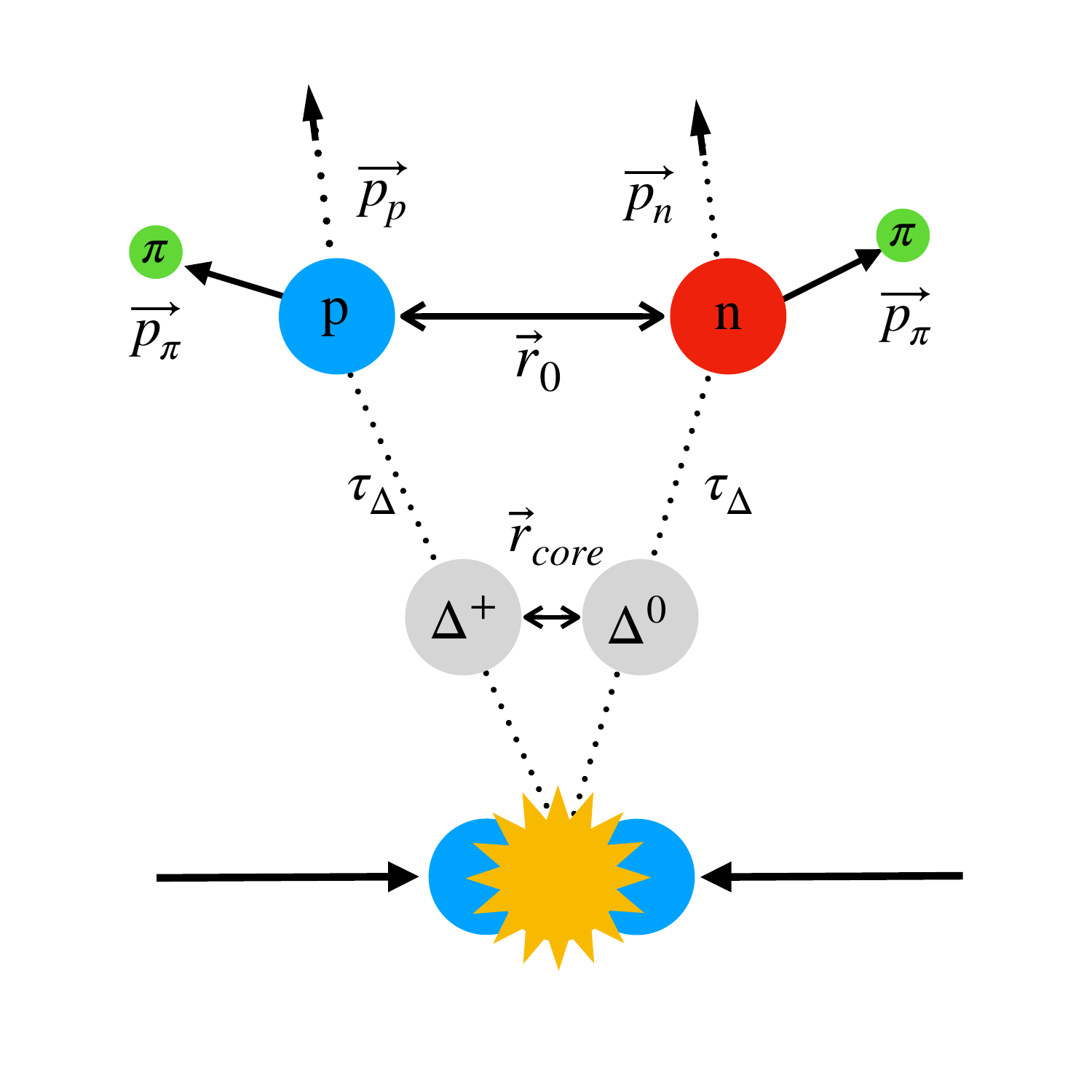}
    \caption{\justifying Pictorial depiction of the resonance source model to describe the emission of (anti-)nucleon pairs in $pp$ collisions.}
    \label{fig:ressrcmodel}
\end{figure}

\subsection{Emission source}\label{emissionsource}
ALICE carries out measurements in estimating the size of nucleon emission sources for $pp$ collisions at \sqs 13 TeV~\cite{ALICE:2020ibs}. These measurements are performed with femtoscopic methods, where the initial spatial correlations are estimated with the help of two-particle momentum correlations. Experimentally, this is obtained via the correlation function for a pair of nucleons having a relative momentum $q^*$ in the pair rest frame, such that
\begin{equation}
    C(q^*) = \frac{1}{N}\frac{A(q^*)}{B(q^*)}
    \label{Cexp}
\end{equation}
$A(q^*)$ and $B(q^*)$ are the two-particle correlation distributions in the same and mixed event, respectively, and $N$ is a normalization constant. The source $S(\mathbf{r})$ can be estimated as
\begin{equation}
    C(\mathbf{q}) = \int d^3r S(\mathbf{r}) \mid\Psi(\mathbf{r},\mathbf{q})\mid^{2}
    \label{Ctheo}
\end{equation}

The source radius $r_{0}$, as described in Eq.~\ref{src}, can be obtained from the fit of Eq.~\ref{Ctheo}. The source function $S(r)$ is radially symmetric, and the relevant one-dimensional (1D) probability density function of $r$ can be written as
\begin{equation}
    S_{4\pi}(r) = 4\pi r^2 S(r) = \frac{4\pi r^2}{(4\pi r_{0}^2)^{3/2}}e^{-\frac{r^2}{4r_{0}^2}}
\end{equation}
\begin{figure}
    \centering
    \includegraphics[scale = 0.45]{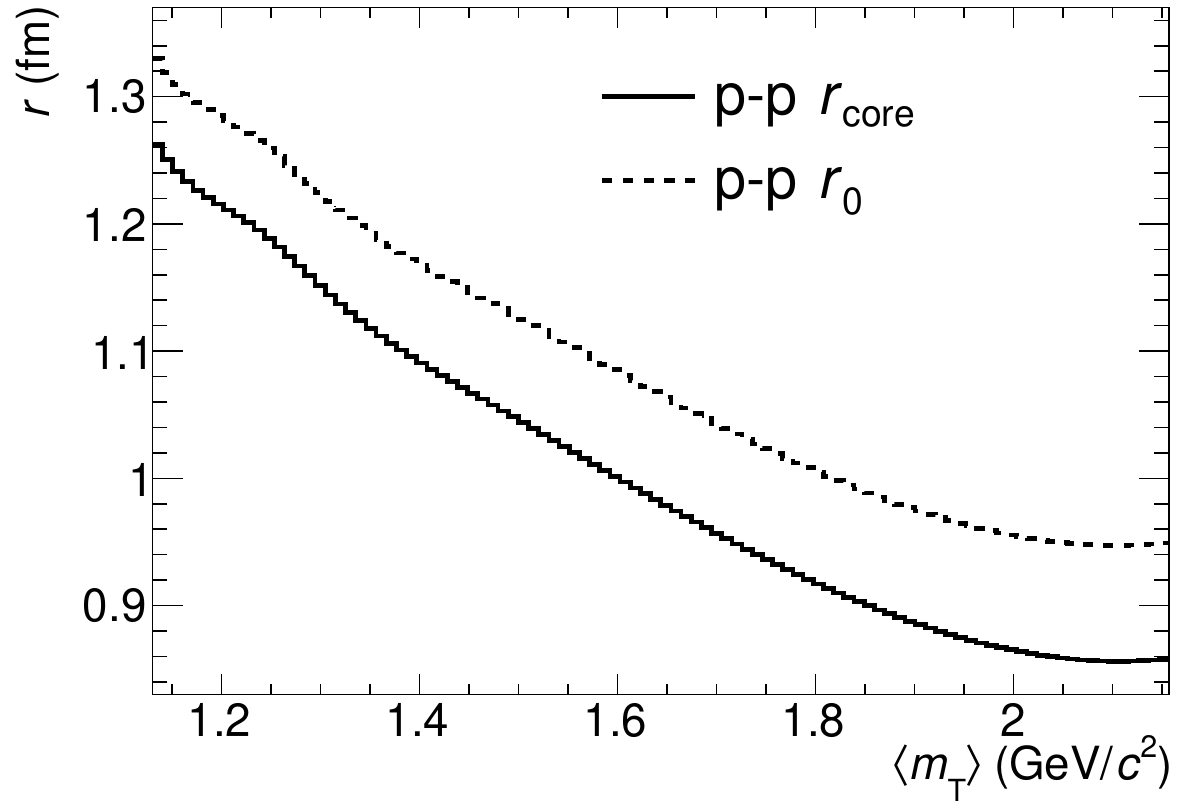}
    \caption{\justifying Source radius ($r_{0}$ and $r_{\rm{core}}$) as a function of pair transverse mass ($\langle m_{\rm{T}} \rangle$) extracted from ALICE measurements~\cite{ALICE:2020ibs}.}
    \label{fig:rmt}
\end{figure}
ALICE measurements have extracted these parameters extensively for baryon systems, namely for $p$ -- $p$ and $p$ -- $\Lambda$ in $pp$ collisions at \sqs 13 TeV. These measurements conclude that the source size vary with (i) the pair transverse mass $\langle m_{\rm{T}} \rangle = \sqrt{ \left(\frac{p_{\mathrm{T}p} + p_{\mathrm{T}n}}{2}\right)^{2} + \left(\frac{m_{p} + m_{n}}{2}\right)^{2}}$ and (ii) according to the decay topology of the baryons. For example, a pair of protons coming from the collision (primordial) and another pair of protons coming from the decay of resonances must have different source radii. This resonance source model assumes that the primordial nucleons and resonances are emitted at equal times, independently from a $``$core'' Gaussian source. The resonances are assumed to be free streaming and non-interacting for their short lifetime. Fig.~\ref{fig:ressrcmodel} shows a pictorial representation of the decay topologies of a proton and neutron pair from resonances. The source radii $r_{\rm{core}}$ belong to the primordial nucleon or resonance pairs coming from the collision, and $r_{0}$ represent the source radii of nucleon pairs, where at least one nucleon is from a resonance decay.\\
\begin{figure}
    \centering
    \includegraphics[scale = 0.4]{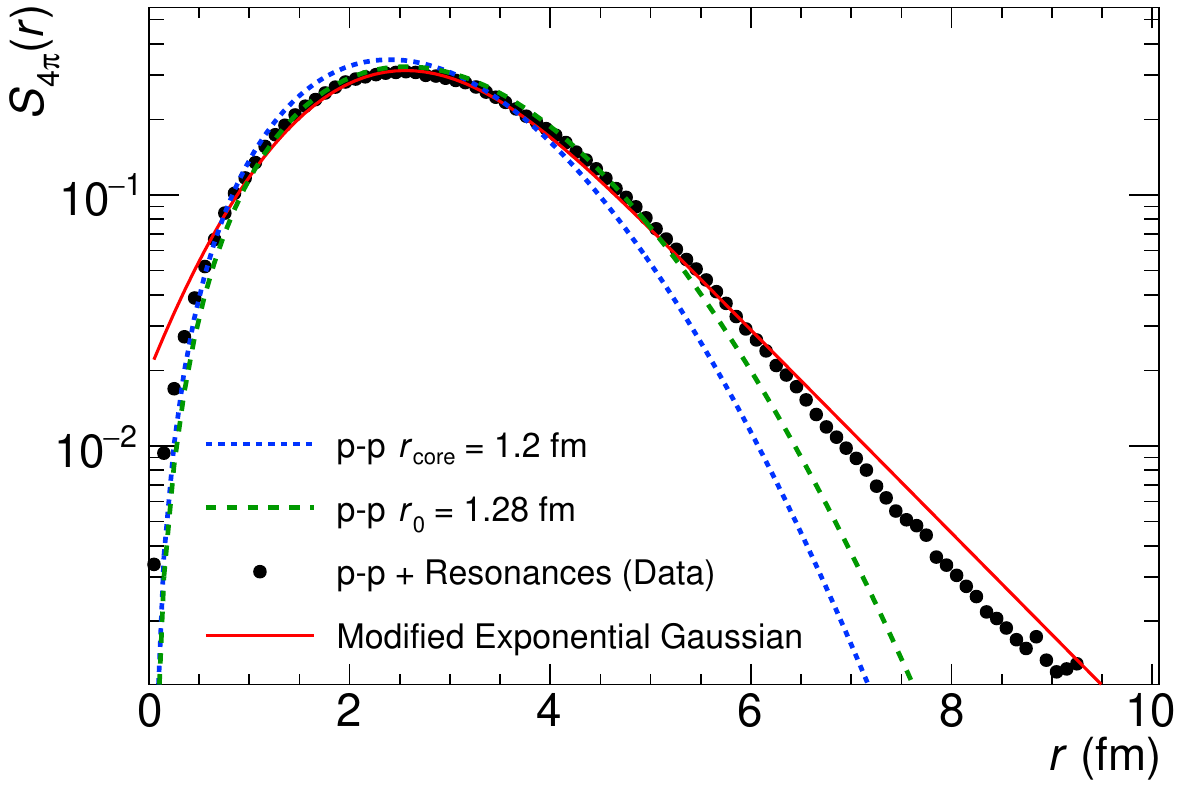}
    \caption{\justifying Source functions for $p$ -- $p$ for primordial $r_{\mathrm{core}}$ = 1.2 fm and from decay emission $r_{0}$ = 1.28 fm. The markers are extracted from the ALICE measurement that shows $p$ -- $p$ sources with one proton coming from a resonance decay. The points are fitted using a modified exponential Gaussian function (solid red line)~\cite{ALICE:2020ibs}.}
    \label{fig:sources}
\end{figure}
Conducting a femtoscopic study to estimate the source sizes with \textsc{pythia}8 is desirable; however, certain discrepancies exist. In \textsc{pythia}8, the source radius of a $p$ -- $p$ state can be calculated from the final state momentum correlations with femtoscopic techniques, which in the $\langle m_\mathrm{T} \rangle$ range 1.26--1.38 GeV/$c$, is estimated to be $\sim$1.2 fm ($r_{\rm{core}}$). It is also seen in the results from reference~\cite{Mahlein:2023fmx} that the source size from \textsc{pythia}8 is relatively non-changing with the $\langle m_\mathrm{T} \rangle$, on contrary to ALICE results, which has a decreasing trend of $r_{\rm{core}}~(r_{0})$ with $\langle m_\mathrm{T} \rangle$. The dynamic nature of $r_{\rm{core}}~(r_{0})$ is important to the coalescence formalism as well as to the likelihood of coalescing in-jet nucleons. The source radii from the ALICE measurements are relied upon, to not overlook this microscopic detail. Fig.~\ref{fig:rmt} shows the source radii values as a function of $\langle m_\mathrm{T} \rangle$. The source radius of a primordial emission ($r_{\rm{core}}$) is more compact than the source with the inclusion of resonance decays. The resonance source model is developed by assigning the emission source radii according to the decay topology of the nucleon pairs and $\langle m_{\rm{T}} \rangle$ as shown in Fig.~\ref{fig:rmt}. This treatment also reinstates the spatiomomenta correlations broken in the factorization of $W_{np}$ into independent spatial and momentum components as shown in Eq.~\ref{wnpfactor}. In Fig.~\ref{fig:sources}, an example of the two sources are presented; the primordial nucleon/resonance pairs having $r_{\rm{core}} = $ 1.2 fm, and a corresponding $r_{0}$ from resonance decays for $\langle m_\mathrm{T} \rangle$ = 1.25 GeV/$c^2$. With the inclusion of the resonances, the Gaussian form is modified by an exponential tail, as seen from the ALICE measurements (markers) in Fig.~\ref{fig:sources}. This shape can be described by a modified Gaussian distribution of the form
\begin{equation}
    f(r;\mu, \sigma, \lambda) = \frac{\lambda}{2}e^{\frac{\lambda}{2}(2\mu + \lambda\sigma^2 - 2r)}\mathrm{erfc}\left( \frac{\mu + \lambda \sigma^2 - r}{\sqrt{2}\sigma}\right)
\end{equation}
where the fit parameters, $\mu = \sqrt{2}r_{0}$ and $\sigma$ is calculated from the Gaussian sources associated to $r_{0}$. The decay parameter $\lambda$ = 0.9 (fixed value) is important to describe the signature tail of the modified source and can be connected to the decay time of the resonance particle. 
\begin{figure}[!ht]
    \centering
    \includegraphics[scale=0.4]{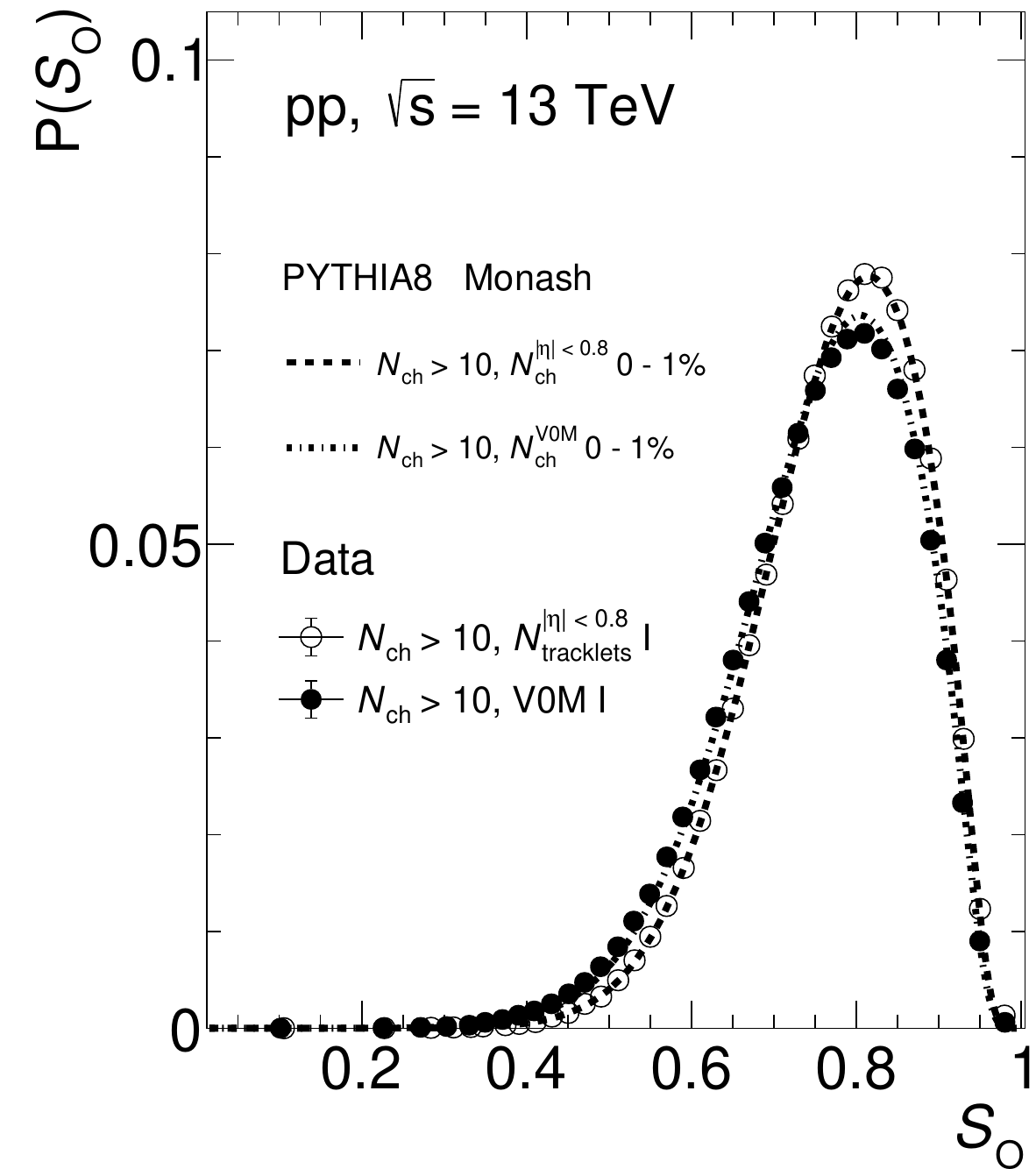}
    \caption{\justifying Transverse spherocity distributions from \textsc{pythia}8 + Monash (lines) for $pp$ collisions at \sqs 13 TeV. Data points from ALICE (markers) in the 0--1\% multiplicity for the midrapidity estimator ($N_\mathrm{tracklet}^{|\eta|<0.8}$ I) and the V0M detector (V0M I) with events having $N_{\rm{ch}} > 10$ are presented for comparison~\cite{ALICE:2023bga}.}
    \label{fig:spherodist}
\end{figure}

\begin{figure*}[!ht]
    \centering
    \includegraphics[scale =0.9]{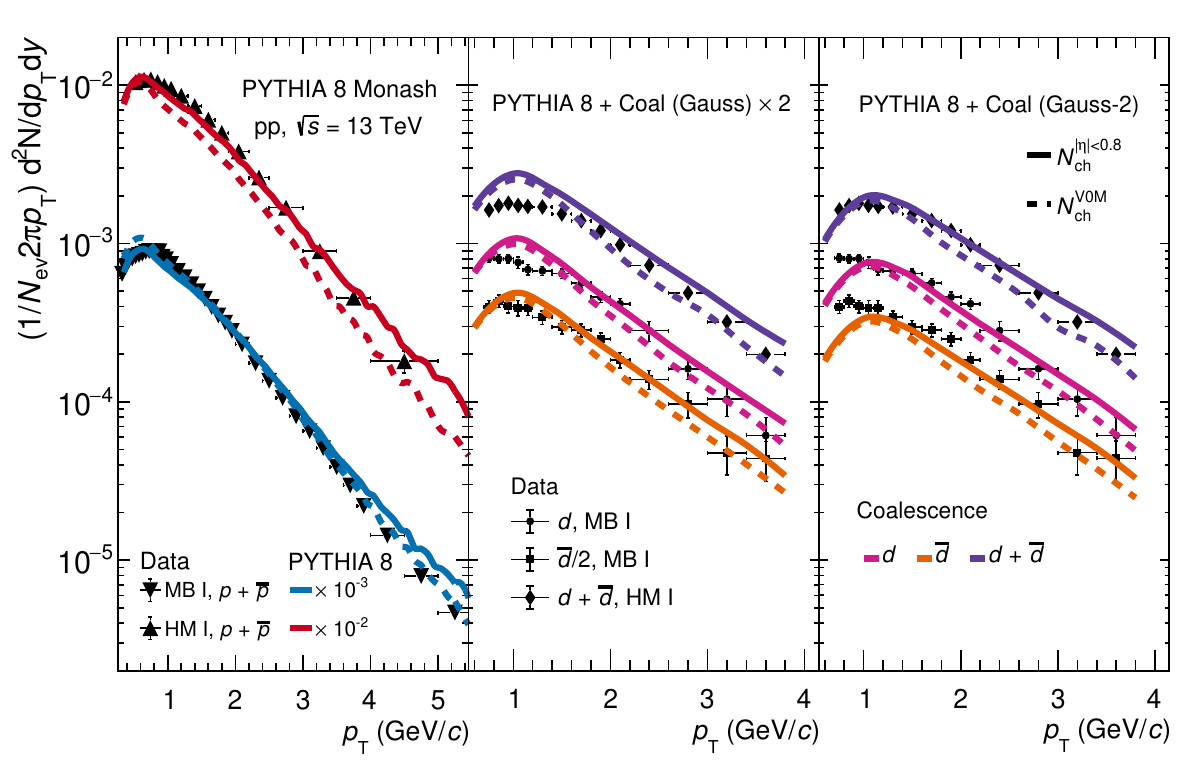}
    \caption{\justifying Transverse momentum spectra of (anti-)protons from \textsc{pythia}8 (left), and deuterons and anti-deuterons from $pp$ collisions at \sqs 13 TeV with \textsc{pythia}8 + Coalescence in MB I and HM I class at midrapidity ($|y| < 0.5$) in two detector acceptances. (Center) Predictions from a single Gaussian deuteron wave function scaled twice to their actual values. (Right) Predictions from a two-Gaussian deuteron wave function. Results are compared to measurements from ALICE in respective multiplicity intervals~\cite{ALICE:2020foi, ALICE:2021mfm}.}
    \label{fig:spectraGaus}
\end{figure*}

\subsection{Transverse spherocity}\label{spherocity}
Event shape observables measure the energy flow deviation in events by classifying the event's structure into a jetty-like or an isotropic one. It is defined in terms of the geometrical distribution of the charged hadrons in the final state and is analytically written as
\begin{equation}
    S_{\rm{O}} = \frac{\pi^2}{4}\min_{\vec{n}=({n_{x},n_{y},0})}\left( \frac{\sum_{i}|\vec{p}_{\rm{T}_{i}}\times\hat{n}|}{\sum_{i}p_{\rm{T}_{i}}}\right)^2
\end{equation}
In order to remove the neutral \pt bias, the \sph is modified to a $``$\pt-unweighted'' form in the manner,
\begin{equation}
    S_{\rm{O}}^{p_{\rm{T}}=1.0} = \frac{\pi^2}{4}\min_{\vec{n}=({n_{x},n_{y},0})}\left( \frac{\sum_{i}|\hat{p}_{\rm{T}_{i}}|_{p_{\rm{T}}=1.0}\times\hat{n}|}{N_{\rm{tracks}}}\right)^2
\end{equation}
where only the angular component of the tracks is in play~\cite{ALICE:2023bga}. For the rest of the paper, $S_{\rm{O}}^{p_{\rm{T}}=1.0}$ is referred to as \sph.\\

Classification of events based on their jettiness allows the investigation of the contributions of hard and soft QCD processes in particle production. Utilizing \sph to search for coalescing nucleons in the jet-likelihood is a reasonable choice. Although \sph does not offer a similar classification to direct \inj and \oj classification, it is close enough to investigate deuterons in events with jets and perform a multi-differential study. In Fig.~\ref{fig:spherodist}, we present the \sph distribution from \textsc{pythia}8 Monash tune, compared to ALICE measurements for 0-1\% minimum bias (MB I) multiplicity in $N_{\rm{tracklet}}$ and V0M acceptances~\cite{ALICE:2023bga}. The events with at least ten tracks are chosen to calculate \sph. \textsc{pythia}8 with the Monash tune describes the \sph distribution for both acceptance ranges in the respective multiplicity classes.\\

\begin{table}
\caption{\justifying \sph selection for \jty and \iso class on different multiplicity classes at midrapidity.}
    \centering
    \begin{tabular}{|c|c|c|} \hline 
         $N_{\rm{ch}}^{\mid \eta \mid < 0.8}$&  Jetty (0--20\%)& Isotropic (80--100\%)\\ \hline 
         (MB I) 0--1\%&  0.665& 0.851\\ \hline 
         (HM I) 0--0.17\%&  0.689& 0.862\\ \hline
    \end{tabular}
    \label{tab:multsphtable}
\end{table}
This work aims to look at coalescence probability in jets, which is performed with two extreme classes of spherocity. The events are divided into percentiles in \sph: 0--20\% (jetty) and 80--100\% (isotropic). The results are presented in MB I and a high multiplicity 0--0.17\% (HM I) class. The \sph intervals used for each multiplicity class at midrapidity are reported in Table~\ref{tab:multsphtable}.

\section{Results and Discussion}
\label{results}

The transverse momentum (\pt) distribution of (anti-)protons and (anti-)deuterons via coalescence are presented in Fig.~\ref{fig:spectraGaus}, for MB I and HM I multiplicity classes in $pp$ collisions at \sqs 13 TeV at midrapidity ($|y| < 0.5$). The (anti-)protons from \textsc{pythia}8 in the MB I class reproduce the data quantitatively, showing good agreement with the V0M estimator and slightly overestimating for the midrapidity estimator for \pt $>$4 GeV/$c$. The HM I (anti-)protons from \textsc{pythia}8 are slightly underestimated by the V0M multiplicity selection but quantitatively captured by the midrapidity estimator, showing a similar overestimation as MB I.\\

The coalescence model predictions with the single Gaussian wave function shown in Fig.~\ref{fig:spectraGaus} (center) are underestimated when compared to experimental measurements from ALICE and are scaled twice to their actual values for better visualisation~\cite{ALICE:2020foi, ALICE:2021mfm}. The predictions for the MB I deuterons and anti-deuterons show a deviation from the experimental measurements at low \pt and slightly underestimates at intermediate to high \pt. Both multiplicity estimators provide a reasonable description of the \pt spectra of the deuterons. The (anti-)deuterons from the HM I sample are also compared to the coalescence model predictions. The disagreement at low \pt is largely visible here, with a reasonable description of the shape from intermediate to high \pt. In this case, the (anti-)deuterons from $N_{\rm{ch}}^{|\eta|<0.8}$ multiplicity sample show a slight overestimation at intermediate to high \pt.\\

The right panel of Fig.~\ref{fig:spectraGaus} presents the model predictions employing the two-Gaussian function as the deuteron wave function. The invariant yields as a function of \pt are nicely described for \pt $>$1.25 GeV/$c$ and deviate at low \pt, similar to the single Gaussian case for MB I and HM I. However, the two-Gaussian model provides a quantitative estimation of the yields. The results from the midrapidity multiplicity estimator show a reasonable agreement with the ALICE measurements~\cite{ALICE:2021mfm}. It is worth noting that through the nature of coalescence, the shapes of the (anti-)deuteron distributions inherit the characteristics of the (anti-)proton, or, more precisely, the (anti-)nucleon momentum distribution. In addition, the distinct shape predicted by the model at low \pt is an artifact of the Wigner probability density associated with the deuteron wave function and the (anti-)nucleon momenta. Moreover, the inconsistencies seen between the multiplicity estimators and the multiplicity classes for the (anti-)proton production from \textsc{pythia}8 also translate to the (anti-)deuteron distribution.\\

The description of the (anti-)deuteron \pt spectra is an important test of the adopted coalescence formalism with its underlying assumptions and the microscopic details of the emission source model. The deuteron wave function is relevant in this problem; the two-Gaussian form fitted to the Hulthen wave function gives a quantitative estimation of the \pt-differential yields over the single Gaussian form. A proper combination of the source model and the light nuclei wave function provides a cohesive description of the coalescence mechanisms of these nuclear clusters.\\

\begin{figure*}[!ht]
    \centering
    \includegraphics[scale = 0.7]{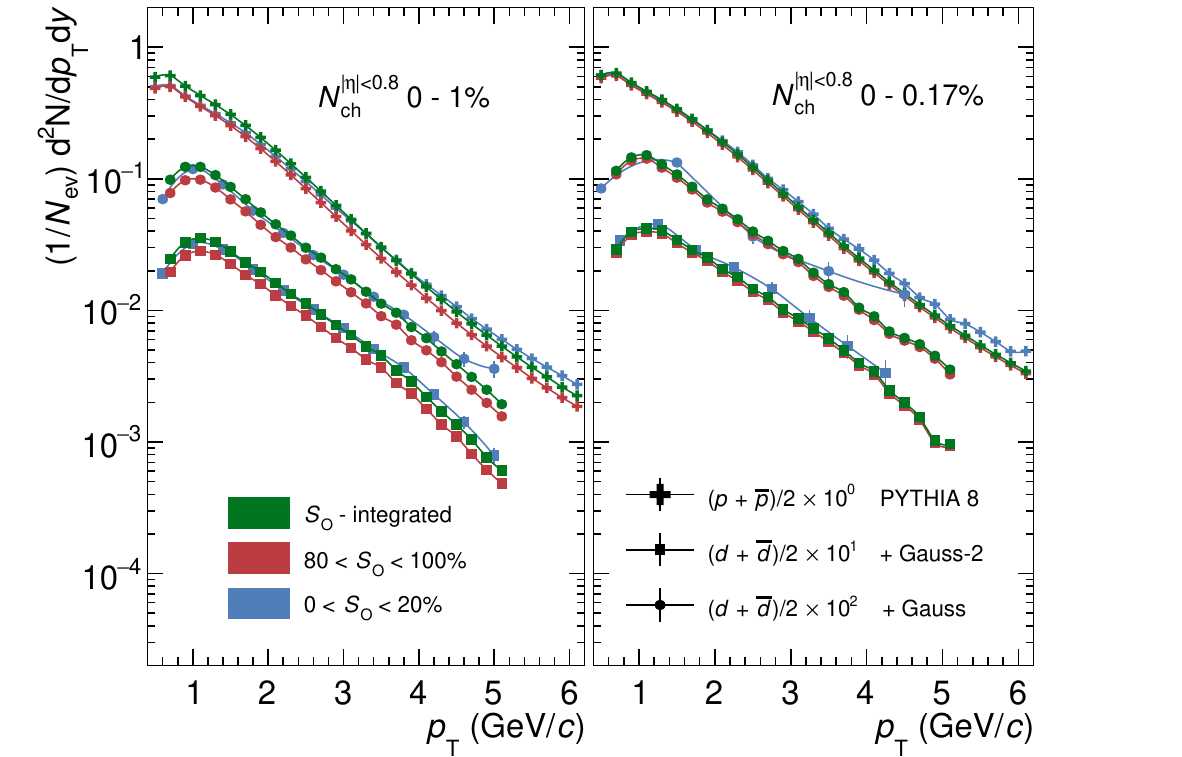}
    \caption{\justifying Transverse momentum spectra of (anti)-protons with \textsc{pythia}8 and (anti-)deuterons with \textsc{pythia}8 + Coalescence in \jty, \iso, and \sph-integrated intervals at midrapidity ($|y| < 0.5$) in $pp$ collisions at \sqs 13 TeV. Predictions from the model are presented in MB I (left) and HM I multiplicity interval (right).}
    \label{fig:SoMultSpectra}
\end{figure*}

We extract the (anti-)deuteron yields at midrapidity into different classes of \sph and $N_{\rm{ch}}^{|\eta|<0.8}$ as described in Table~\ref{tab:multsphtable}. The \pt spectra of (anti-)protons and (anti-)deuterons from each of these selections are shown in Fig.~\ref{fig:SoMultSpectra} for the single and double Gaussian wave functions. The \jty (\iso) (anti-)protons and (anti-)deuterons show hardening (softening) of the spectra at high (low) \pt when compared to the \sph integrated \pt spectra at MB I multiplicity. The \jty (anti-)deuterons in HM I are affected by statistical uncertainties at intermediate to high \pt, which makes the same observation inconclusive. It is clear that for separate \sph intervals, the (anti-)deuteron yields share the behavior of (anti-)proton yields. We do not note any peculiarity in the (anti-)deuteron yields. These trends present the nature of the production of protons and deuterons in different event topologies; the harder event carries more protons at high-\pt and vice versa.\\

\begin{figure*}[!ht]
    \centering
    \includegraphics[scale = 0.7]{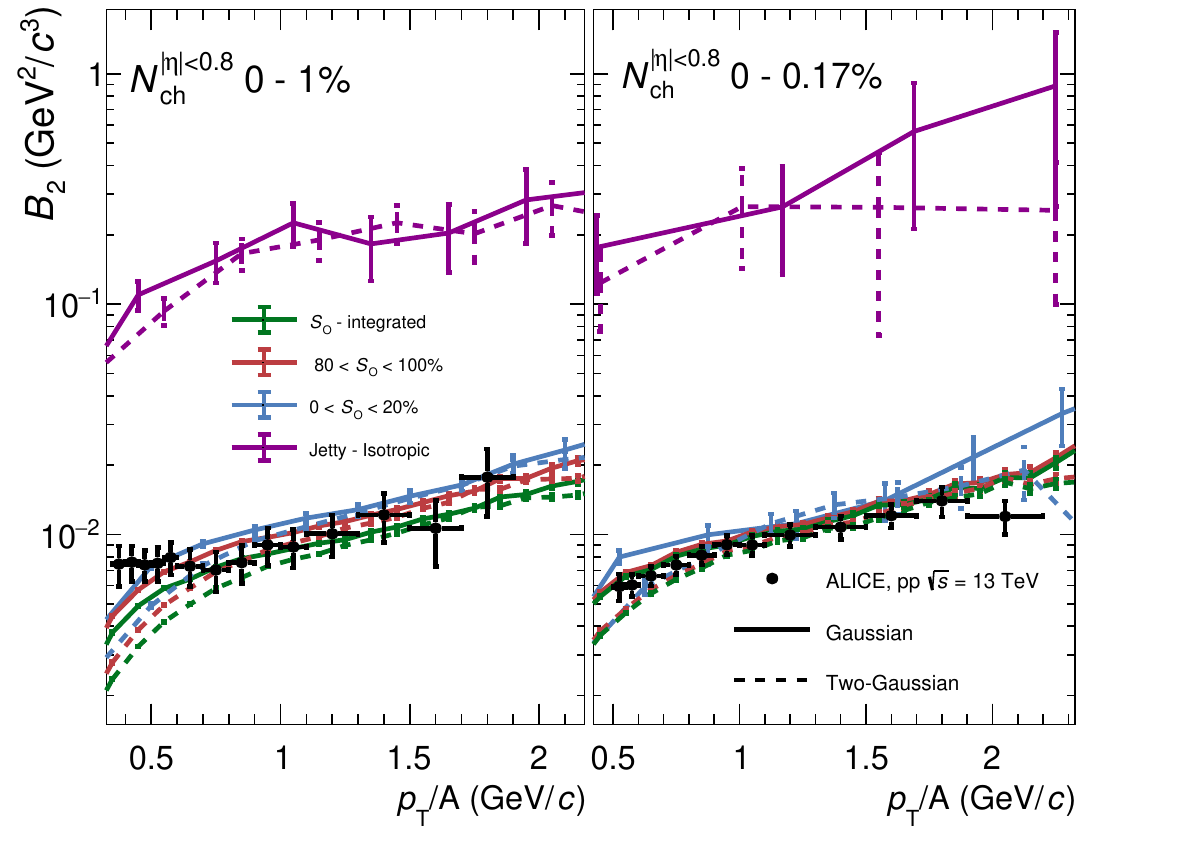}
    \caption{\justifying Coalescence parameter of (anti-)deuterons $B_2$ as a function of proton transverse momentum at midrapidity ($|y| < 0.5$) in $pp$ collisions at \sqs 13 TeV. Solid and dashed lines represent the predictions from single and double Gaussian wave functions respectively. The results are compared to experimental measurements from ALICE~\cite{ALICE:2020foi, ALICE:2021mfm}.}
    \label{fig:coalfac}
\end{figure*}

In Fig.~\ref{fig:coalfac}, the coalescence parameter, $B_2$ as a function of \pt at midrapidity, is presented. The coalescence parameter is calculated using Eq.~\ref{eqnb2} and shown for 0--20\%, 80--100\%, and integrated \sph in MB I and HM I multiplicities. In both multiplicity intervals, the \jty deuterons show a slightly higher $B_2$, which increases with increasing \pt. Although noticeable, the difference between the three \sph cases is insignificant. The quoted ALICE measurements belong to the \sph-integrated interval, which is supported by the calculations from the model~\cite{ALICE:2020foi, ALICE:2021mfm}. These results also show that the \jty, \iso, and the integrated case are close or comparable, which can be accredited to the contribution of the UE at high multiplicities. The UE plays a dominant role in deuteron production.\\

The degree of enhancement of the $B_2$ for \jty deuterons is not as significant as the one reported by ALICE, where the \inj deuterons show a $B_2$ 10 times more than those of the UE~\cite{ALICE:2022ugx}. In the ALICE measurement, the \inj deuterons (region towards the jet) are separated by subtracting the UE contributions (region away from the jet) from the (anti-)deuterons closer to the jet. This removes the contribution of the UE deuterons in the region towards the jet and keeps the deuterons that originate solely from the jet fragmentation. This procedure can be repeated for the deuterons in \sph intervals by subtracting the \iso deuterons from the \jty ones. These isolated (anti-)deuterons are produced from the coalescence of correlated (anti-)nucleon pairs that are produced in the jet fragmentation processes in events that carry a \jty topology. However, one must be cautious as the deuterons would belong to different events from being truly separated by \sph without knowing the actual contribution of the UE in a \sph class. To approximate this, the average number of MPIs ($\langle N_{\rm{MPI}}\rangle$) is calculated for 0--20\% and 80--100\% \sph intervals. The relative contribution of $\langle N_{\rm{MPI}}\rangle$ in \jty to \iso events is taken as a weight for the \iso (anti-)deuteron \pt distribution. The \inj deuterons are then approximated by taking the difference of \jty and the weighted \iso distribution. At high multiplicities, where the contribution of MPI is large, there is a small difference between the number of MPIs between a jetty and an isotropic event. The relative contribution from MPIs in \jty to \iso events is calculated to be $\approx$90--95\%\\

Fig.~\ref{fig:coalfac} shows the (anti-)deuteron coalescence parameter for $``$jetty-isotropic'' events. In both multiplicity intervals, the $B_2$ of \jty-\iso deuterons is much higher than the individual \sph classes, showing an apparent $``$enhancement''. The predictions from both models are compatible with one another. The \jty - \iso $B_2$ is ten times to the \sph integrated $B_2$ for the MB I multiplicity and up to 25 times for the HM I multiplicity class. However, the statistical uncertainties for the HM I \inj deuterons are too large to draw a conclusive estimate of the enhancement. Although a quantitative comparison with ALICE measurements cannot be made as the multiplicity selections are different, the observations made on the enhancement of the coalescence probability in this study are quite transparent. With this enhancement in $B_2$, we also observe that the strong spatiomomenta correlations between the coalescing nucleons are restored by removing the \iso or UE deuterons from the \jty deuterons that contain an amalgamation of deuterons from both UE and jet fragmentation.\\

The results on (anti-)deuteron $B_2$ corroborate the claim on $B_2$ enhancement for \inj deuterons and the observations on the same performed by ALICE~\cite{ALICE:2022ugx}. Although the separation of \inj deuterons is performed in an approximate manner, the motivation behind this work is supported by the results showing a clear enhancement of $B_2$ for \inj deuterons compared to the ones belonging to the UE. It also shows that (anti-)deuteron production is dominated by the UE in $pp$ collisions. Additionally, a finite \pt dependence of $B_2$ is observed in the MB I multiplicity class results, portraying a similar trend to the spectra from the UE. The same cannot be concluded for the HM class due to high statistical uncertainties.\\

\section{Conclusions}\label{conclusions}

In this article, we studied the production mechanisms of (anti-)deuteron in $pp$ collisions at midrapidity with an advanced coalescence model and tested it for high multiplicity $pp$ collisions and different event shape intervals. The (anti-)deuteron coalescence model is based on the Wigner probability density of a deuteron wave function. The coalescence mechanism is supplemented by the microscopic details of the emission source size in $pp$ collisions. Using the results from ALICE femtoscopic correlations of baryon sources, the relative distances between the nucleons as a function of transverse mass are used as spatial inputs for the model. Moreover, the decay of resonances and their contribution to the production of nucleons are also considered, which presents an interesting complication to the emission model. The sources are controlled by the nature of the decay topology of the nucleon, which is a Gaussian when both nucleons belong to a $``$core'' or a decaying (anti-)nucleon emission. A modified Gaussian is considered for a mixed case, which fits well with the calculations. The modified Gaussian inherits the properties of the pure Gaussian source, adding a fixed decay parameter $\lambda$ that captures the decay time of the resonances.\\

The resonance-source model is embedded into the \textsc{pythia}8 event generator, where selected nucleon pairs are assigned relative distances. Based on the relative distances and momenta of the nucleons, the Wigner probability is calculated. Two choices of the deuteron wave function are presented; a single Gaussian and a double Gaussian form parameterized to fit the deuteron Hulthen wave function. The (anti-)deuteron yields at midrapidity from the single Gaussian wave function underestimate the experimental result and provide a qualitative estimate of the yields. The two-Gaussian form gives a quantitative estimate of the yields, predicting the experimental measurements in both multiplicity intervals. The (anti-)deuteron distributions are influenced by the Wigner probability density used and the (anti-)nucleon momenta. The coalescence model also inherits the characteristics of the nucleon distribution, which was noted from the similar discrepancies observed between (anti-)proton and (anti-)deuterons in different multiplicities and multiplicity estimators.\\

To search for hints of enhancement of coalescence of deuterons in jets, an event-shape differential measurement is performed using transverse spherocity. The deuteron production is investigated in \jty and \iso spherocity classes in two multiplicity intervals. The \jty (\iso) \pt-differential yields of (anti-)protons and (anti-)deuterons show hardness (softness) concerning the \sph-integrated spectra. The coalescence parameter $B_2$ is estimated in each \sph interval, and the values are comparable. The \jty deuterons show a slight increase in $B_2$, although insignificant. This is accredited to the UE's significant contribution at high multiplicities, sharing a larger contingent of the (anti-)deuteron yield.\\

To look deeper into the \inj deuterons, the contribution of the UE is subtracted from the deuterons of \jty events. The isotropic deuterons, purely from the UE, serve as a proxy. The fraction of MPI in \jty and \iso events is taken to estimate the UE. The $``$jetty-isotropic'' $B_2$ show a clear enhancement compared to the individual \sph spectra. The $``$jetty-isotropic'' deuterons serve as a good approximation of the \inj deuterons. The degree of enhancement of $B_2$ can be compared to the ALICE measurements, which cross ten times for \inj deuterons to the deuterons from the UE. The enhancement of $B_2$ is due to the favorable coalescence conditions put forward by the restoration of the strong spatiomomenta correlations of the nucleon pairs produced from the jet fragmentation.\\

This study emphasizes the importance of an advanced coalescence model, which comes with the unified formalism of the deuteron wave function and an emission source model. Further studies with the application of state-of-the-art deuteron wave functions will provide further insight into the production and dynamics of the deuteron via coalescence. With precise femtoscopy studies, one can dive deep into the coalescence mechanism and constrain the effects of \inj and the UE in (anti-)deuteron in $pp$ collisions. 

\section{Acknowledgments} 
Y.B. thanks all the authors of \textsc{pythia}8. Y.B. is grateful to Sudhir P. Rode for carefully reading the manuscript and to Sumit Kundu for contributing to the data-generating process. Y.B. is also thankful to Ravindra Singh and Swapnesh Khade for the fruitful discussions. This work uses computational facilities supported by the DST-FIST scheme via SERB Grant No. SR/FST/PSI-225/2016, by the Department of Science and Technology (DST), Government of India.
%\bibliography{bibliography}




\begin{thebibliography}{50}
%\cite{British-Scandinavian-MIT:1977tan}
\bibitem{British-Scandinavian-MIT:1977tan}
S.~Henning \textit{et al.} [British-Scandinavian-MIT],
%``Production of Deuterons and anti-Deuterons in Proton Proton Collisions at the CERN ISR,''
Lett. Nuovo Cim. \textbf{21} (1978), 189
doi:10.1007/BF02822248
%56 citations counted in INSPIRE as of 10 Mar 2022

%\cite{Alper:1973my}
\bibitem{Alper:1973my}
B.~Alper, H.~B\ensuremath{\gimel}gild, P.~Booth, F.~Bulos, L.~J.~Carroll, G.~von Dardel, G.~Damgaard, B.~Duff, F.~Heymann and J.~N.~Jackson, \textit{et al.}
%``Large angle production of stable particles heavier than the proton and a search for quarks at the cern intersecting storage rings,''
Phys. Lett. B \textbf{46} (1973), 265-268
doi:10.1016/0370-2693(73)90700-4
%98 citations counted in INSPIRE as of 10 Mar 2022

%\cite{E878:1998vna}
\bibitem{E878:1998vna}
M.~J.~Bennett \textit{et al.} [E878],
%``Light nuclei production in relativistic Au + nucleus collisions,''
Phys. Rev. C \textbf{58} (1998), 1155-1164
doi:10.1103/PhysRevC.58.1155
%29 citations counted in INSPIRE as of 18 Dec 2023

%\cite{E802:1999hit}
\bibitem{E802:1999hit}
L.~Ahle \textit{et al.} [E802],
%``Proton and deuteron production in Au + Au reactions at 11.6/A-GeV/c,''
Phys. Rev. C \textbf{60} (1999), 064901
doi:10.1103/PhysRevC.60.064901
%147 citations counted in INSPIRE as of 18 Dec 2023

%\cite{STAR:2016ydv}
\bibitem{STAR:2016ydv}
L.~Adamczyk \textit{et al.} [STAR],
%``Measurement of elliptic flow of light nuclei at $\sqrt{s_{NN}}=$ 200, 62.4, 39, 27, 19.6, 11.5, and 7.7 GeV at the BNL Relativistic Heavy Ion Collider,''
Phys. Rev. C \textbf{94} (2016) no.3, 034908
doi:10.1103/PhysRevC.94.034908
[arXiv:1601.07052 [nucl-ex]].
%77 citations counted in INSPIRE as of 18 Dec 2023

%\cite{STAR:2019sjh}
\bibitem{STAR:2019sjh}
J.~Adam \textit{et al.} [STAR],
%``Beam energy dependence of (anti-)deuteron production in Au + Au collisions at the BNL Relativistic Heavy Ion Collider,''
Phys. Rev. C \textbf{99} (2019) no.6, 064905
doi:10.1103/PhysRevC.99.064905
[arXiv:1903.11778 [nucl-ex]].
%89 citations counted in INSPIRE as of 18 Dec 2023

%\cite{ALICE:2015wav}
\bibitem{ALICE:2015wav}
J.~Adam \textit{et al.} [ALICE],
%``Production of light nuclei and anti-nuclei in pp and Pb-Pb collisions at energies available at the CERN Large Hadron Collider,''
Phys. Rev. C \textbf{93} (2016) no.2, 024917
doi:10.1103/PhysRevC.93.024917
[arXiv:1506.08951 [nucl-ex]].
%255 citations counted in INSPIRE as of 18 Dec 2023

%\cite{ALICE:2017xrp}
\bibitem{ALICE:2017xrp}
S.~Acharya \textit{et al.} [ALICE],
%``Production of deuterons, tritons, $^{3}$He nuclei and their antinuclei in pp collisions at $\mathbf{\sqrt{{\textit s}}}$ = 0.9, 2.76 and 7 TeV,''
Phys. Rev. C \textbf{97} (2018) no.2, 024615
doi:10.1103/PhysRevC.97.024615
[arXiv:1709.08522 [nucl-ex]].
%122 citations counted in INSPIRE as of 18 Dec 2023

%\cite{ALICE:2020foi}
\bibitem{ALICE:2020foi}
S.~Acharya \textit{et al.} [ALICE],
%``(Anti-)deuteron production in pp collisions at $\sqrt{s}=13 \ \text {TeV}$,''
Eur. Phys. J. C \textbf{80} (2020) no.9, 889
doi:10.1140/epjc/s10052-020-8256-4
[arXiv:2003.03184 [nucl-ex]].
%42 citations counted in INSPIRE as of 23 Dec 2022

%\cite{ALICE:2021mfm}
\bibitem{ALICE:2021mfm}
S.~Acharya \textit{et al.} [ALICE],
%``Production of light (anti)nuclei in pp collisions at $ \sqrt{s} $ = 13 TeV,''
JHEP \textbf{01} (2022), 106
doi:10.1007/JHEP01(2022)106
[arXiv:2109.13026 [nucl-ex]].
%23 citations counted in INSPIRE as of 26 Oct 2023

%\cite{ALICE:2017jmf}
\bibitem{ALICE:2017jmf}
S.~Acharya \textit{et al.} [ALICE],
%``Production of $^{4}$He and $^{4}\overline{\textrm{He}}$ in Pb-Pb collisions at $\sqrt{s_{\mathrm{NN}}}$ = 2.76 TeV at the LHC,''
Nucl. Phys. A \textbf{971} (2018), 1-20
doi:10.1016/j.nuclphysa.2017.12.004
[arXiv:1710.07531 [nucl-ex]].
%112 citations counted in INSPIRE as of 18 Dec 2023

%\cite{STAR:2010gyg}
\bibitem{STAR:2010gyg}
B.~I.~Abelev \textit{et al.} [STAR],
%``Observation of an Antimatter Hypernucleus,''
Science \textbf{328} (2010), 58-62
doi:10.1126/science.1183980
[arXiv:1003.2030 [nucl-ex]].
%258 citations counted in INSPIRE as of 22 Dec 2022

%\cite{Winkler:2020ltd}
\bibitem{Winkler:2020ltd}
M.~W.~Winkler and T.~Linden,
%``Dark Matter Annihilation Can Produce a Detectable Antihelium Flux through $\bar{\Lambda}_b$ Decays,''
Phys. Rev. Lett. \textbf{126} (2021) no.10, 101101
doi:10.1103/PhysRevLett.126.101101
[arXiv:2006.16251 [hep-ph]].
%18 citations counted in INSPIRE as of 23 Dec 2022

%\cite{ALICE:2022zuz}
\bibitem{ALICE:2022zuz}
S.~Acharya \textit{et al.} [ALICE],
%``Measurement of anti-$^3$He nuclei absorption in matter and impact on their propagation in the Galaxy.,''
doi:10.1038/s41567-022-01804-8
[arXiv:2202.01549 [nucl-ex]].
%6 citations counted in INSPIRE as of 23 Dec 2022

%\cite{Butler:1963pp}
\bibitem{Butler:1963pp}
S.~T.~Butler and C.~A.~Pearson,
%``Deuterons from High-Energy Proton Bombardment of Matter,''
Phys. Rev. \textbf{129} (1963), 836-842
doi:10.1103/PhysRev.129.836
%310 citations counted in INSPIRE as of 18 Dec 2023

%\cite{Kapusta:1980zz}
\bibitem{Kapusta:1980zz}
J.~I.~Kapusta,
%``Mechanisms for deuteron production in relativistic nuclear collisions,''
Phys. Rev. C \textbf{21} (1980), 1301-1310
doi:10.1103/PhysRevC.21.1301
%178 citations counted in INSPIRE as of 18 Dec 2023

%\cite{Scheibl:1998tk}
\bibitem{Scheibl:1998tk}
R.~Scheibl and U.~W.~Heinz,
%``Coalescence and flow in ultrarelativistic heavy ion collisions,''
Phys. Rev. C \textbf{59} (1999), 1585-1602
doi:10.1103/PhysRevC.59.1585
[arXiv:nucl-th/9809092 [nucl-th]].
%267 citations counted in INSPIRE as of 18 Dec 2023

%\cite{Zhao:2018lyf}
\bibitem{Zhao:2018lyf}
W.~Zhao, L.~Zhu, H.~Zheng, C.~M.~Ko and H.~Song,
%``Spectra and flow of light nuclei in relativistic heavy ion collisions at energies available at the BNL Relativistic Heavy Ion Collider and at the CERN Large Hadron Collider,''
Phys. Rev. C \textbf{98} (2018) no.5, 054905
doi:10.1103/PhysRevC.98.054905
[arXiv:1807.02813 [nucl-th]].
%55 citations counted in INSPIRE as of 18 Dec 2023

%\cite{Sun:2018mqq}
\bibitem{Sun:2018mqq}
K.~J.~Sun, C.~M.~Ko and B.~D\"onigus,
%``Suppression of light nuclei production in collisions of small systems at the Large Hadron Collider,''
Phys. Lett. B \textbf{792} (2019), 132-137
doi:10.1016/j.physletb.2019.03.033
[arXiv:1812.05175 [nucl-th]].
%84 citations counted in INSPIRE as of 18 Dec 2023

%\cite{Steinheimer:2012tb}
\bibitem{Steinheimer:2012tb}
J.~Steinheimer, K.~Gudima, A.~Botvina, I.~Mishustin, M.~Bleicher and H.~Stocker,
%``Hypernuclei, dibaryon and antinuclei production in high energy heavy ion collisions: Thermal production versus Coalescence,''
Phys. Lett. B \textbf{714} (2012), 85-91
doi:10.1016/j.physletb.2012.06.069
[arXiv:1203.2547 [nucl-th]].
%123 citations counted in INSPIRE as of 10 Mar 2022

%\cite{Andronic:2010qu}
\bibitem{Andronic:2010qu}
A.~Andronic, P.~Braun-Munzinger, J.~Stachel and H.~Stocker,
%``Production of light nuclei, hypernuclei and their antiparticles in relativistic nuclear collisions,''
Phys. Lett. B \textbf{697} (2011), 203-207
doi:10.1016/j.physletb.2011.01.053
[arXiv:1010.2995 [nucl-th]].
%334 citations counted in INSPIRE as of 23 Dec 2022

%\cite{Becattini:2014hla}
\bibitem{Becattini:2014hla}
F.~Becattini, E.~Grossi, M.~Bleicher, J.~Steinheimer and R.~Stock,
%``Centrality dependence of hadronization and chemical freeze-out conditions in heavy ion collisions at $\sqrt s_{NN}$ = 2.76 TeV,''
Phys. Rev. C \textbf{90} (2014) no.5, 054907
doi:10.1103/PhysRevC.90.054907
[arXiv:1405.0710 [nucl-th]].
%89 citations counted in INSPIRE as of 18 Dec 2023

%\cite{Vovchenko:2018fiy}
\bibitem{Vovchenko:2018fiy}
V.~Vovchenko, B.~D\"onigus and H.~Stoecker,
%``Multiplicity dependence of light nuclei production at LHC energies in the canonical statistical model,''
Phys. Lett. B \textbf{785} (2018), 171-174
doi:10.1016/j.physletb.2018.08.041
[arXiv:1808.05245 [hep-ph]].
%75 citations counted in INSPIRE as of 18 Dec 2023

%\cite{Kachelriess:2023jis}
\bibitem{Kachelriess:2023jis}
M.~Kachelriess, S.~Ostapchenko and J.~Tjemsland,
%``Effect of nonequal emission times and space-time correlations on (anti-) nuclei production,''
Phys. Rev. C \textbf{108} (2023) no.2, 024903
doi:10.1103/PhysRevC.108.024903
[arXiv:2303.08437 [hep-ph]].
%0 citations counted in INSPIRE as of 26 Oct 2023

%\cite{ALICE:2022ugx}
\bibitem{ALICE:2022ugx}
S.~Acharya \textit{et al.} [ALICE],
%``Enhanced Deuteron Coalescence Probability in Jets,''
Phys. Rev. Lett. \textbf{131} (2023) no.4, 042301
doi:10.1103/PhysRevLett.131.042301
[arXiv:2211.15204 [nucl-ex]].
%3 citations counted in INSPIRE as of 26 Oct 2023


%\cite{ALICE:2020hjy}
\bibitem{ALICE:2020hjy}
S.~Acharya \textit{et al.} [ALICE],
%``Jet-associated deuteron production in pp collisions at $\sqrt{s}$ = 13 TeV,''
Phys. Lett. B \textbf{819} (2021), 136440
doi:10.1016/j.physletb.2021.136440
[arXiv:2011.05898 [nucl-ex]].
%12 citations counted in INSPIRE as of 26 Oct 2023

%\cite{ALICE:2017qfj}
\bibitem{ALICE:2017qfj}
 [ALICE],
%``Supplemental material: afterburner for generating light (anti-)nuclei with QCD-inspired event generators in pp collisions,''
ALICE-PUBLIC-2017-010.
%5 citations counted in INSPIRE as of 23 Dec 2022

%\cite{JETSCAPE:2022cob}
\bibitem{JETSCAPE:2022cob}
D.~Everett \textit{et al.} [JETSCAPE],
%``Role of bulk viscosity in deuteron production in ultrarelativistic nuclear collisions,''
Phys. Rev. C \textbf{106} (2022) no.6, 064901
doi:10.1103/PhysRevC.106.064901
[arXiv:2203.08286 [hep-ph]].
%6 citations counted in INSPIRE as of 26 Oct 2023

%\cite{Kachelriess:2019taq}
\bibitem{Kachelriess:2019taq}
M.~Kachelrie\ss{}, S.~Ostapchenko and J.~Tjemsland,
%``Alternative coalescence model for deuteron, tritium, helium-3 and their antinuclei,''
Eur. Phys. J. A \textbf{56} (2020) no.1, 4
doi:10.1140/epja/s10050-019-00007-9
[arXiv:1905.01192 [hep-ph]].
%26 citations counted in INSPIRE as of 26 Oct 2023

%\cite{Kachelriess:2020amp}
\bibitem{Kachelriess:2020amp}
M.~Kachelriess, S.~Ostapchenko and J.~Tjemsland,
%``On nuclear coalescence in small interacting systems,''
Eur. Phys. J. A \textbf{57} (2021) no.5, 167
doi:10.1140/epja/s10050-021-00469-w
[arXiv:2012.04352 [hep-ph]].
%12 citations counted in INSPIRE as of 26 Oct 2023

%\cite{ALICE:2020ibs}
\bibitem{ALICE:2020ibs}
S.~Acharya \textit{et al.} [ALICE],
%``Search for a common baryon source in high-multiplicity pp collisions at the LHC,''
Phys. Lett. B \textbf{811} (2020), 135849
doi:10.1016/j.physletb.2020.135849
[arXiv:2004.08018 [nucl-ex]].
%64 citations counted in INSPIRE as of 26 Oct 2023

%\cite{Mahlein:2023fmx}
\bibitem{Mahlein:2023fmx}
M.~Mahlein, L.~Barioglio, F.~Bellini, L.~Fabbietti, C.~Pinto, B.~Singh and S.~Tripathy,
%``A realistic coalescence model for deuteron production,''
Eur. Phys. J. C \textbf{83} (2023) no.9, 804
doi:10.1140/epjc/s10052-023-11972-3
[arXiv:2302.12696 [hep-ex]].
%3 citations counted in INSPIRE as of 26 Oct 2023

%\cite{Sjostrand:2000wi}
\bibitem{Sjostrand:2000wi}
T.~Sjostrand, P.~Eden, C.~Friberg, L.~Lonnblad, G.~Miu, S.~Mrenna and E.~Norrbin,
%``High-energy physics event generation with PYTHIA 6.1,''
Comput. Phys. Commun. \textbf{135} (2001), 238-259
doi:10.1016/S0010-4655(00)00236-8
[arXiv:hep-ph/0010017 [hep-ph]].
%3308 citations counted in INSPIRE as of 26 Oct 2023

%\cite{Sjostrand:2007gs}
\bibitem{Sjostrand:2007gs}
T.~Sjostrand, S.~Mrenna and P.~Z.~Skands,
%``A Brief Introduction to PYTHIA 8.1,''
Comput. Phys. Commun. \textbf{178} (2008), 852-867
doi:10.1016/j.cpc.2008.01.036
[arXiv:0710.3820 [hep-ph]].
%7186 citations counted in INSPIRE as of 26 Oct 2023

%\cite{Bierlich:2022pfr}
\bibitem{Bierlich:2022pfr}
C.~Bierlich, S.~Chakraborty, N.~Desai, L.~Gellersen, I.~Helenius, P.~Ilten, L.~L\"onnblad, S.~Mrenna, S.~Prestel and C.~T.~Preuss, \textit{et al.}
%``A comprehensive guide to the physics and usage of PYTHIA 8.3,''
doi:10.21468/SciPostPhysCodeb.8
[arXiv:2203.11601 [hep-ph]].
%300 citations counted in INSPIRE as of 26 Oct 2023

%\cite{Skands:2014pea}
\bibitem{Skands:2014pea}
P.~Skands, S.~Carrazza and J.~Rojo,
%``Tuning PYTHIA 8.1: the Monash 2013 Tune,''
Eur. Phys. J. C \textbf{74} (2014) no.8, 3024
doi:10.1140/epjc/s10052-014-3024-y
[arXiv:1404.5630 [hep-ph]].
%1207 citations counted in INSPIRE as of 18 Dec 2023

%\cite{Sjostrand:1987su}
\bibitem{Sjostrand:1987su}
T.~Sjostrand and M.~van Zijl,
%``A Multiple Interaction Model for the Event Structure in Hadron Collisions,''
Phys. Rev. D \textbf{36} (1987), 2019
doi:10.1103/PhysRevD.36.2019
%1051 citations counted in INSPIRE as of 26 Oct 2023


%\cite{Argyropoulos:2014zoa}
\bibitem{Argyropoulos:2014zoa}
S.~Argyropoulos and T.~Sj\"ostrand,
%``Effects of color reconnection on $t\bar{t}$ final states at the LHC,''
JHEP \textbf{11} (2014), 043
doi:10.1007/JHEP11(2014)043
[arXiv:1407.6653 [hep-ph]].
%164 citations counted in INSPIRE as of 26 Oct 2023

%\cite{Sjostrand:2004pf}
\bibitem{Sjostrand:2004pf}
T.~Sjostrand and P.~Z.~Skands,
%``Multiple interactions and the structure of beam remnants,''
JHEP \textbf{03} (2004), 053
doi:10.1088/1126-6708/2004/03/053
[arXiv:hep-ph/0402078 [hep-ph]].
%313 citations counted in INSPIRE as of 26 Oct 2023


%\cite{Andersson:1983ia}
\bibitem{Andersson:1983ia}
B.~Andersson, G.~Gustafson, G.~Ingelman and T.~Sjostrand,
%``Parton Fragmentation and String Dynamics,''
Phys. Rept. \textbf{97} (1983), 31-145
doi:10.1016/0370-1573(83)90080-7
%4021 citations counted in INSPIRE as of 26 Oct 2023

%\cite{ALICE:2020swj}
\bibitem{ALICE:2020swj}
S.~Acharya \textit{et al.} [ALICE],
%``Pseudorapidity distributions of charged particles as a function of mid- and forward rapidity multiplicities in pp collisions at $\sqrt{s}$~=~5.02, 7 and 13 TeV,''
Eur. Phys. J. C \textbf{81} (2021) no.7, 630
doi:10.1140/epjc/s10052-021-09349-5
[arXiv:2009.09434 [nucl-ex]].
%33 citations counted in INSPIRE as of 26 Oct 2023

%\cite{ALICE:2019avo}
\bibitem{ALICE:2019avo}
S.~Acharya \textit{et al.} [ALICE],
%``Multiplicity dependence of (multi-)strange hadron production in proton-proton collisions at $\sqrt{s}$ = 13 TeV,''
Eur. Phys. J. C \textbf{80} (2020) no.2, 167
doi:10.1140/epjc/s10052-020-7673-8
[arXiv:1908.01861 [nucl-ex]].
%106 citations counted in INSPIRE as of 18 Dec 2023

%\cite{Bellini:2018epz}
\bibitem{Bellini:2018epz}
F.~Bellini and A.~P.~Kalweit,
%``Testing production scenarios for (anti-)(hyper-)nuclei and exotica at energies available at the CERN Large Hadron Collider,''
Phys. Rev. C \textbf{99}, no.5, 054905 (2019)
doi:10.1103/PhysRevC.99.054905
[arXiv:1807.05894 [hep-ph]].
%73 citations counted in INSPIRE as of 01 Feb 2024

%\cite{ALICE:2023bga}
\bibitem{ALICE:2023bga}
S.~Acharya \textit{et al.} [ALICE],
%``Light-flavor particle production in high-multiplicity pp collisions at $\mathbf{\sqrt{\textit{s}} = 13}$ TeV as a function of transverse spherocity,''
[arXiv:2310.10236 [hep-ex]].
%0 citations counted in INSPIRE as of 26 Oct 2023




\end{thebibliography}


 \end{document}